\def\<{\langle}
\def\>{\rangle}

\def\p{{\bf p}}

\def\delp{{\delta p}}

\def\beq{\begin{equation}}
\def\eeq{\end{equation}}

\documentclass[preprint,aps,draft]{revtex4}
\usepackage{amsmath}
\usepackage{amssymb}
\usepackage{amsfonts}
\usepackage{graphics}                                                         

\begin{document}
\title{Relativistic Quantum Theories and Neutrino Oscillations}
\author{B. D. Keister}
\affiliation{
Physics Division, 1015N \\
National Science Foundation \\
4201 Wilson Blvd. \\
Arlington, VA  22230} 

\author{W. N. Polyzou}
\affiliation{ 
Department of Physics and Astronomy \\
The University of Iowa \\ 
Iowa City, IA, 52242
}

\vspace{10mm}

\begin{abstract}

  Neutrino oscillations are examined under the broad requirements of
  Poincar\'e-invariant scattering theory in an $S$-matrix formulation.
  This approach can be consistently applied to theories with either
  field or particle degrees of freedom.  The goal of this paper is to
  use this general framework to identify all of the unique physical
  properties of this problem that lead to a simple oscillation
  formula.  We discuss what is in principle observable, and how many
  factors that are important in principle end up being negligible in
  practice.

\end{abstract}

\maketitle

\section{Introduction}

\label{sec:intro}

There is now strong experimental evidence that neutrino flavor
eigenstates are mixed via a non-diagonal matrix that connects them to
neutrino mass eigenstates~\cite{AnnualReview:2008,overview}.  The fact that space-time
propagation is governed by mass rather than flavor eigenstates gives
rise to the possibility of oscillation between flavor states over
space and time.

There is a large body of literature addressing various aspects of the
quantum mechanics of neutrino oscillation.  Many of the primary issues
were set forth by Kayser~\cite{Kayser:1981} in 1981.  Since then, this
subject has been examined from a variety of
perspectives~\cite{Giunti:1991, Rich:1993, Grimus:1996,
Stodolsky:1998, Beuthe:2002, Beuthe:2003, Giunti:2003,
Lipkin:2004, Cohen:2008qb, Akhmedov:2009rb}.

The phenemenon of neutrino oscillations seems simple enough from the
perspective of introductory quantum mechanics, for which there are
countless examples of oscillations in two- and three-level systems.
There are, however, differences in the neutrino case that tend to work
against the intuition of the standard examples: the initial conditions
for neutrino production in reactors or the Sun are generally not
controlled; the coherence length over which states are virtual can be
kilometers rather than subatomic distances, and this
affects considerations such as what is large or small in a
calculation; one of the weak interactions that determine the
oscillation interval takes place inside the volume of the detector
rather than far from it; the neutrino kinematics are
ultrarelativistic, but not fully so.  There is also the distinction
between what is measured in an experiment and features of a specific
theoretical approach.  These issues have led to a variety of
approaches as cited above.  Some of these approaches have differing
perspectives~\cite{Giunti:2003,Lipkin:2004}, yet all lead to the same
simple oscillation formula.

In this paper, we provide an approach to neutrino oscillations from
the perspective of scattering theory in relativistic quantum
mechanics.  The virtue of this approach is its generality and its
focus on what can actually be measured in oscillation experiments.  The
scattering theory approach advocated in this paper has been
examined~\cite{Cardall:2000} using quantum field theory, but the
conclusions that follow from our approach are not limited to a field
theoretic treatment.  There are many papers on neutrino oscillations,
and many of them have significant overlap with this work.  Our goal
here is to provide a very general framework utilizing the $S$ matrix.

The $S$ matrix for a neutrino oscillation experiment is the
probability amplitude for transitions between states where both
initial and final particles are localized near two macroscopically
separated space-time points; the points of neutrino creation and
neutrino absorption.  We discuss the additional physical conditions
that must be satisfied for the transition amplitude which, when
squared, leads to a neutrino oscillation formula.  There is also a
freedom in the choice of variables used to label the 
single-particle intermediate states, the choice of scattering equivalent
Hamiltonian, as well as the choice of kinematic symmetries of the
interaction.  Different choices lead to the same scattering matrix,
but will normally lead to different formulae in specific
implementations and different interpretations of the dynamics.  We
find that the conditions that lead to a simple oscillation formula
also combine to render as insignificant many factors that might
otherwise serve to distinguish among various theoretical approaches.

\section{Formal Basis: Poincar\'e Invariant Quantum Mechanics}
\label{sec:rqm}

We present here an overview of the basic ingredients of quantum
mechanical systems that satisfy relativistic invariance.  Further
details can be found in Ref.~\cite{Review:1990}.

Relativistic invariance in a quantum theory means that a change of
inertial coordinate system is a symmetry of the theory.  In special
relativity, inertial coordinate systems are related by space-time
translations and proper orthochronous Lorentz transformations.  The
group generated by these transformations is the Poincar\'e group.
Proper orthochronous Lorentz transformations do not include the
discrete transformations associated space reflection and time
reversal, which are broken by the weak interaction.  Wigner proved
\cite{Wigner:1939} that the existence of a unitary representation,
$U(\Lambda ,a)$ of the Poincar\'e group is both a necessary and a
sufficient condition for a quantum theory to be relativistically
invariant.

The dynamics are given by the time-translation subgroup of the
Poincar\'e group, which is generated by the Hamiltonian, $H$.  
Consistency of the initial value problem requires that
interactions must also appear in additional Poincar\'e generators.
This means that at most a subgroup of the Poincar\'e group can be
independent of interactions.  When such a subgroup exists, it is
called a kinematic subgroup.  While the existence of a kinematic
subgroup is a choice of representation,
the choice of representation may imply kinematic
symmetries of the interaction that have no impact on the scattering
matrix elements; this is an example of a representation-dependent
feature of an interaction that is not experimentally observable.

The largest kinematic subgroups were classified by Dirac
\cite{Dirac:1949}; the classification was completed by Patera and
Winternitz \cite{Patera:1975}. The three largest kinematic subgroups
are the three-dimensional Euclidean group (instant-form dynamics), the
Lorentz group (point-form dynamics) and the subgroup that leaves a
three-dimensional hyper-plane tangent to the light cone invariant
(null-plane dynamics).

Interactions in a dynamical model where the kinematic subgroup is the
three-dimensional Euclidean group are translationally invariant and
thus conserve three-momentum; interactions in a dynamical model with a
null-plane kinematic symmetry are invariant with respect to
translations in the null plane and thus conserve components of the
four momentum that the generate translations on the null plane; while
interactions in a dynamical model where the kinematic symmetry is the
Lorentz group are Lorentz invariant.
 
Scattering theory in a Poincar\'e invariant quantum theory can be
formulated using the same time-dependent methods, based on dynamical
and asymptotic Hamiltonians, $H$ and $H_0$.  that are used in both
non-relativistic quantum mechanics and quantum field
theory.

Scattering states $\vert \Psi (t) \rangle$ are solutions of the 
time-evolution equation (Schr\"odinger equation)
\beq 
\vert \Psi_{\pm} (t) \rangle = e^{-iHt} \vert \Psi_{\pm} (0) \rangle,
\label{a.0}
\eeq
where the initial condition, $\vert \Psi_{\pm} (0) \rangle$, is 
determined by an asymptotic
condition.  There are two natural scattering asymptotic conditions;
they require that the state $ \vert \Psi_{\pm} (t) \rangle$ approach a
state of non-interacting particles, $\vert \Phi_{\pm} (t) \rangle $,
in the asymptotic future or past:
\beq
\lim_{t \to \pm \infty} \Vert \vert \Psi_{\pm}  (t) \rangle
- \vert \Phi_{\pm} (t) \rangle \Vert = 0
\label{a.1}
\eeq
where 
\beq
\vert \Phi_{\pm} (t) \rangle
=e^{-iH_0t} \vert \Phi_{\pm} (0) \rangle
\label{a.1a}
\eeq
is the non-interacting state.  In the field theoretic case
the $H_0$ must include self-interactions.  The appropriate generalization 
of the asymptotic condition (\ref{a.1}), first formulated by Haag and Ruelle,
is discussed in \cite{brenig}\cite{simon}\cite{jost}.

Equations (\ref{a.0}-\ref{a.1a}) imply that the initial condition 
for the scattering states
are related to the initial conditions for the non-interacting states by
\beq
\vert \Psi_{\pm}  (0) \rangle = \lim_{t \to \pm \infty} 
e^{iHt} e^{-iH_0t} \vert \Phi_{\pm} (0) \rangle :=
\Omega_{\pm} \vert \Phi_{\pm} (0) \rangle .
\label{a.1b}
\eeq

The probability amplitude that a state prepared to become
$\vert \Phi_- (t) \rangle$ in the asymptotic past scatters into a
state that becomes $\vert \Phi_+ (t) \rangle$ in the asymptotic
future is
\beq
\langle S \rangle  = \langle \Psi_{+}  (t) \vert \Psi_{-}  (t) \rangle =
\langle \Psi_{+}  (0) \vert \Psi_{-}  (0) \rangle =
\langle \Phi_{+} (0) \vert
S
\vert \Phi_{-} (0) \rangle 
\eeq
where
\beq
S:= \Omega^{\dagger}_{+} \Omega_{-}
\eeq
is the scattering operator, and we have exploited the invariance of 
the probability amplitudes under
time translation to emphasize that the probability amplitude can be
computed using states at any common time.  We will use this in formulating
scattering involving neutrino intermediate states.

We note that if $H$ is transformed with a unitary
transformation $A$ such that $H'=AHA^{\dagger}$, and $A$ satisfies
\beq
\label{A-asym}
\lim_{t \to \pm \infty} \Vert (I-A)e^{-iH_0t} \vert \Psi \rangle \Vert =0  
\eeq
for {\it both} time limits then 
\beq
S= S (H,H_0)= S(H',H_0).
\eeq 
Note that $H_0$ is not transformed, so the representation of the
asymptotically free particles remains unchanged.  This means $V=H-H_0$
and $V'=H'-H_0$ are distinct interactions that cannot be
experimentally distinguished.  Operators $A$ satisfying the asymptotic
property (\ref{A-asym}) are called scattering equivalences.  They
preserve the scattering matrix without changing $H_0$.
Ekstein \cite{Ekstein:1960} proved that the existence of such
operators is a necessary and a sufficient condition for two
Hamiltonians to be scattering equivalent.

Any experiment that only measures scattering matrix elements cannot
distinguish different scattering equivalent Hamiltonians.  
If a dynamical representation of the Poincar\'e group has a given
kinematic subgroup, it is possible to construct unitary scattering
equivalences \cite{Polyzou:2002} that change the kinematic subgroup to
any other kinematic subgroup, so any properties attributed to a particular 
kinematic symmetry are not observable.

\section{Asymptotic states} 

The probability amplitude for a transition from the prepared initial
state to the measured final state can be expressed in terms of a
scattering operator $S$.
As discussed in the previous section, the scattering probability
amplitude is the inner product of two solutions of the time-evolution
equation with past and future asymptotic conditions evaluated at any
common time.  We can describe neutrino scattering reactions using an
$S$-matrix approach.  This was first done by
Cardall~\cite{Cardall:2000} within a field-theoretic framework.

As an illustrative example we consider a reaction where an initial proton and
electron interact to produce a linear combination of neutrino mass
eigenstates which propagate over a macroscopic distance until they
are absorbed by a second interaction that produces a final proton and
electron.  In the language of second-order perturbation theory this
will involve two successive reactions, e.g.,
\begin{eqnarray}
p + e &\to n' + \nu\, (a)
\nonumber
\\
n + \nu &\to p' + e'\, (b).
\label{eq:b.0}
\end{eqnarray}
This example can be generalized to other cases of interest without
affecting the overall conclusions, since the role of the intermediate
neutrino mass eigenstates is the same.  In this example, the
asymptotic states are given by $\{p,n,e,p',n',e'\}$ and not the
neutrino mass eigenstates, which are virtual.

For the cases of interest, reaction $(b)$ takes place inside a
neutrino detector volume, and another device records that event via
the emerging charged lepton.  This is distinct from more typical
applications in which the $S$ matrix describes one or more events that
are distinctly separated from any detection equipment. 

The asymptotic free-particle state before the reaction is represented by a
localized wave packet describing an electron $e$ moving toward the proton
$p$; in addition, there is also a free neutron $n$ that is
traveling toward the point where it will eventually interact with
the neutrino mass eigenstates.

Similarly, the asymptotic free-particle state after the reaction is
represented by a localized wave packet describing a neutron $n'$
traveling away from the region of the initial interaction, and an electron
$e'$ and proton $p'$ traveling away from the point where the neutrino
mass eigenstate was absorbed by the initial neutron $n$.

There is a definite probability amplitude for a transition from the
initial state containing $\{p,e,n\}$ to the final state containing
$\{p',e',n'\}$.  The unusual feature is that there is not a single
localized space-time region where the initial and final states
overlap.  Because we only need to construct the scattering states,
$\vert \Psi_{\pm} (t) \rangle$, whose inner product can be evaluated
at any common time, the $S$ matrix can be computed in the same manner
that is used with more traditional asymptotic states.

For the example discussed above, the reaction is characterized by two
disjoint space-time regions localized about $x_a$, where the coherent
neutrino mass eigenstate superposition is created, and $x_b$, where it
is absorbed. In our example the initial proton $p$
and electron $e$ and final neutron $n'$ are localized near $x_a$ at a
common time $t_a$, and the initial neutron $n$ and final proton $p'$
and electron $e'$ are localized near $x_b$ at a later common time
$t_b$.

In what follows we generalize the expressions to include all lepton
flavors $(e,\mu,\tau)$, labeled by Greek indices $\alpha, \beta$.

The construction of asymptotic states corresponding to this reaction
starts with normalizable single-particle states that localize the
particles near $\mathbf{x}_a$ at time $t_a$ or $\mathbf{x}_b$ 
at time $t_b$ with the appropriate
expectation value for the initial or final momenta of the observed
particles.  In order to localize these states at the different points
$\mathbf{x}_a$ and $\mathbf{x}_b$ we initially localize them at the
origin and use single particle spatial translations, $T_j(-\mathbf{x}_c) \vert
\phi_j \rangle$, $c=a$ or $b$, to localize the $j$-th particle near
$\mathbf{x}_a$ or $\mathbf{x}_b$. 
The $-$ sign is consistent with 
\begin{equation}
\mathbf{x}_c=
\langle \phi \vert  (\mathbf{x} + \mathbf{x_c} ) 
\vert \phi_j \rangle =
\langle \phi_j \vert T_j(\mathbf{x_c}) \mathbf{x} T_j^{\dagger}(\mathbf{x_c})
\vert \phi_j \rangle =
\langle \phi_j \vert T_j^{\dagger}(-\mathbf{x}_c) 
\mathbf{x} T_j (-\mathbf{x}_c)
\vert \phi \rangle .
\end{equation}

From Eq.~\ref{a.1b}, we have that
\beq
\vert \Psi_{\pm} (t) \rangle = \Omega_{\pm} \vert \Phi_{\pm} (t) \rangle.
\eeq
This means that the initial and final scattering states can be transformed to a 
common time by transforming the non-interacting asymptotic states 
to the common time.
To construct the non-interacting multi-particle asymptotic states at a common
time, all of the particles that are near $\mathbf{x}_b$ at time $t_b$
are time-translated to where they would be at time $t=t_a$ using
single-particle time evolution:
\begin{equation}
T_j(t_a-t_b) T_j(-\mathbf{x}_b) \vert \phi_j \rangle = 
T_j(t_a-t_b,-\mathbf{x}_b) \vert \phi_j \rangle .
\label{eq:b.4}
\end{equation}
Choosing $t_a=0$ gives initial and final time-zero asymptotic states 
of the form
\begin{equation}
\vert \Phi_\alpha (0) \rangle = \prod_i T(-{x}_a)
\vert \phi_{i\alpha}  \rangle  
\prod_j T(-{x}_b) \vert \phi_{j \alpha}  \rangle .
\label{eq:b.5}
\end{equation}
Here the single-particle states $\vert \phi_{i \alpha} \rangle$ are
initially localized at the origin, and all of the ``reaction
geometry'' appears in the one-body space-time translation operators,
$T_i (-x_i)$.  In the absence of interactions these non-interacting
asymptotic states, $\vert \Phi_\alpha (0) \rangle$, put the initial or
final reaction products that interact at time $t=t_a=0$ near
$\mathbf{x}_a$ and the initial and final reaction products that
interact at time $t=t_b$ near $\mathbf{x}_b$.  Because all of the
``reaction geometry'' is in the structure of $\vert \Phi_\alpha
(0) \rangle$, the formal scattering operator, $S$, is unchanged and
can be calculated using standard methods of quantum mechanics or
quantum field theory.

In what follows we use the notation $S_{fi} (x_b,x_a)$ for $S$-matrix
elements with asymptotic states describing neutrinos created at
space-time point $x_a$ and absorbed at space-time point $x_b$.

In this formulation of the scattering problem, 
the space-time displacement, $x_{ba} := x_b-x_a$, transforms
like a four-vector under Lorentz transformations.  To show this, note
that the scattering matrix element is Lorentz invariant:
\begin{equation}
\vert \Psi^\pm_\alpha  \rangle \to 
\vert \Psi^{\pm\prime}_\alpha  \rangle :=  
U(\Lambda ,0) \vert \Psi^\pm_\alpha  \rangle 
\qquad \rightarrow \qquad
S_{fi}=S'_{fi} .
\label{eq:b.6}
\end{equation}
The effect of this transformation on the asymptotic states follows
from the Poincar\'e invariance of the wave operators
\begin{equation}
U(\Lambda,a ) \Omega_{\pm} = \Omega_{\pm} U_0 (\Lambda ,a)    
\label{eq:b.7}
\end{equation}
where $U_0 (\Lambda ,a )$ is a product of
single-particle unitary representations of the 
Poincar\'e group, $U_j (\Lambda ,a )$.   The Poincar\'e invariance 
of the wave operators ensures that the representations of the 
Poincar\'e group for the free and interacting systems agree 
when the particles are asymptotically separated.
Because $T_j(x_c) = U_j(I,x_c)$,  
the group representation property 
implies
\begin{equation}
U_j(\Lambda ,0 )T_j(-x_c) = T_j(-\Lambda x_c  ) U_j(\Lambda ,0 ).
\label{eq:b.8}
\end{equation}
It follows from (\ref{eq:b.7}) and (\ref{eq:b.8}) that 
\begin{equation}
S_{fi} (x_b,x_a) = S_{f'i'} (\Lambda x_a, \Lambda x_b)
\label{eq:b.9}
\end{equation}
where the single particle states localized at the origin are replaced by 
the transformed states localized at the origin
\begin{equation}
\vert \phi'_{j \alpha}  \rangle = U_j (\Lambda ,0) \vert \phi_{j \alpha}  
\rangle .
\label{eq:b.10}
\end{equation}
Thus, in this formalism, dynamical Poincar\'e transformations imply
the expected transformation properties of the parameters that describe
the geometry of the interaction region.  This will be relevant for the
oscillation formula, which in this formalism is determined by
considering how the scattering operator behaves when these parameters
are varied.

\section{ Second-order S matrix}

In this section we examine the conditions that lead to neutrino
oscillations in the $S$-matrix approach.  We assume that the weak
interaction can be treated perturbatively.  In our example the leading
contribution to the scattering matrix is of second order in the weak
interaction.  As noted above, the initial weak interaction at
space-time point $x_a$ produces a superposition of neutrino (or
anti-neutrino) mass eigenstates, which propagate to the space-time
point $x_b$ of the final weak interaction.  We assume that the initial
particle(s) collide or decay near the space-time point $x_a$ producing
a neutrino or anti-neutrino and final particle(s).  The neutrino or
anti-neutrino travels and interacts with initial particle(s) near
space-time point $x_b$ to produce final particles.  The superposition
of neutrino mass eigenstates is never directly observed; it appears
only as an intermediate state.  The second-order calculation can
formally be done in the interaction representation, where the $H_0$
includes all interactions except for the interactions with the
neutrinos.

For the purpose of illustration, we will assume that the kinematic
subgroup of the Poincar\'e group is the three-dimensional Euclidean
group, generated by spatial translations and rotations.  This is
called an ``instant form'' dynamics.  In an instant-form dynamics it
is natural to label single-particle basis vectors,
$\vert \mathbf{p}\rangle$, by the particle's three momentum
$\mathbf{p}$, while the fourth component $p^0$ satisfies the
mass-shell condition $p^0 = (m^2 +
\mathbf{p}^2)^{1\over2}$.  Spin labels are suppressed.  The weak
interaction, $V$, that couples to the neutrinos creates flavor
eigenstates that can be decomposed into mass eigenstates that
propagate.  In an instant-form dynamics the neutrino production
interaction is invariant with respect to kinematic translations and
thus conserves the sum of the single-particle momenta.  
As mentioned earlier, this conservation law
is related to the freedom to choose among scattering-equivalent
interactions and does not affect the $S$ matrix.  There are many
other scattering-equivalent interactions that do not have this 
kinematic symmetry.

The second-order contribution to the kernel of scattering operator, $S$,
for our example reaction is 
\[
\langle \mathbf{p}_{\beta b}', \mathbf{p}_{pb}', \mathbf{p}_{na}' 
\vert S \vert \mathbf{p}_{\alpha a},\mathbf{p}_{pa} ,
\mathbf{p}_{nb}  \rangle=
\]
\[
-2 \pi i 
\delta^4 ( p_a + p_b - p_a' - p_b' )
\times
\]
\begin{equation}
\sum_{j=1}^3  {\langle  (\mathbf{p}_{\beta b}',\mathbf{p}_{pb}')^+ \Vert 
V^{\dagger}_{j\beta} \Vert \mathbf{p}_{nb} , 
\mathbf{p}_{\nu_j} \rangle  
\langle \mathbf{p}_{na}' ,
\mathbf{p}_{\nu_j} \Vert V_{j\alpha} \Vert 
(\mathbf{p}_{\alpha a},  \mathbf{p}_{pa})^- \rangle
\over E_{a} - E_{a'} - E_{\nu_j}+ i 0^+} ,
\label{eq:c.1}
\end{equation}
where $p_a, p_b ,p_a',p_b'$ are the initial and final four momenta at
points $a$ and $b$. The Roman index $j$ labels the mass eigenstate,
and the neutrino energy is 
\beq
E_{\nu_j} = \sqrt{m_{\nu_j} - \mathbf{p}_{\nu_j}^2}.
\label{eq:c.4}
\eeq
In addition, the following relations are
specific to an interaction with the three-dimensional Euclidean 
group as a kinematic subgroup:
\beq
\langle \mathbf{p}_{na}' ,
\mathbf{p}_{\nu_j} \vert V_{j\alpha} \vert 
(\mathbf{p}_{\alpha a},  \mathbf{p}_{pa})^- \rangle =
\delta (\mathbf{p}_{na}' +
\mathbf{p}_{\nu_j} - 
\mathbf{p}_{\alpha a}-  \mathbf{p}_{pa})
\langle \mathbf{p}_{na}' ,
\mathbf{p}_{\nu_j} \Vert V_{j\alpha} \Vert 
(\mathbf{p}_{\alpha a},  \mathbf{p}_{pa})^- \rangle .
\label{c.4.a}
\eeq

The sum in (\ref{eq:c.1}) is over neutrino mass eigenstates.  The
$()^{\pm}$ states are electron-proton scattering eigenstates.
Equation (\ref{eq:c.1}) is a generalization of the standard
``two-potential'' formula of Gell-Mann and Goldberger
\cite{Gellman:1953} where cluster properties have been used to
factorize the incoming and outgoing scattering states into products of
independent scattering states associated with the reactions at $x_a$
and $x_b$.
 
The neutrinos enter in the matrix elements
\begin{equation}
\langle  (\mathbf{p}_{\beta b}',\mathbf{p}_{pb}')^+ \Vert 
V^{\dagger}_{j\beta} \Vert \mathbf{p}_{nb} , 
\mathbf{p}_{\nu_j} \rangle
\end{equation}
and 
\begin{equation}  
\langle \mathbf{p}_{na}' ,
\mathbf{p}_{\nu_j} \Vert V_{j\alpha} \Vert 
(\mathbf{p}_{\alpha a}, \mathbf{p}_{pa})^- \rangle . 
\end{equation} 
The three-momentum delta function in Eq.~(\ref{c.4.a}) implies that 
in this representation the
virtual neutrino three-momentum is constrained by the external kinematics:
\begin{equation}  
\mathbf{p}_{\nu_j} =
\mathbf{p}_{\alpha a}-  \mathbf{p}_{pa} - \mathbf{p}_{na} .
\label{3-momentum}
\end{equation} 

The flavor of the initial and final states determines the flavor of the 
created/detected neutrinos.  These matrix elements are constructed from 
an elementary vertex that involves the neutrino mixing matrix.

It is useful to include explicitly the one-body space-time translation
operators used in the construction of the non-interacting 
asymptotic states with the
kernel of the scattering operator, $S$.  Using the four-momentum conservation of the
second-order contribution to the $S$ operator gives the expression
\[
e^{i (p_{\beta b}'+ p_{p b}' - p_{n b})\cdot (x_b-x_a)} 
\langle \mathbf{p}_{\beta b}', \mathbf{p}_{pb}', \mathbf{p}_{na}'   
\vert S \vert \mathbf{p}_{\alpha a}, \mathbf{p}_{pa} ,
\mathbf{p}_{nb}  \rangle=
\]
\[
-2 \pi i 
\delta^4 ( p_{\alpha a} + p_{pa} + p_{nb} - p_{\beta b}' - p_{pb}' - p_{na}'
) e^{i (p_{\beta b}' + p_{pb}'- p_{nb}) \cdot (x_b-x_a)}
\times
\]
\begin{equation}
\sum_j  {
\langle  (\mathbf{p}_{\beta b}',\mathbf{p}_{pb}')^+ \Vert 
V^{\dagger}_{j\beta} \Vert \mathbf{p}_{nb} , 
\mathbf{p}_{\nu_j} \rangle
\langle \mathbf{p}_{na}' ,
\mathbf{p}_{\nu_j} \Vert V_{j\alpha} \Vert 
(\mathbf{p}_{\alpha a},  \mathbf{p}_{pa})^- \rangle 
\over E_{\alpha a} + E_{pa} - E_{na}' - E_{\nu_j} + i0^+} . 
\label{eq:c.9}
\end{equation}

A probability amplitude is obtained by integrating this kernel over
initial and final wave packets associated with states localized at the
origin.  The relevant matrix elements have the form
\[
S_{fi}(x_b,x_a) =
\]
\[
\int 
\phi_{\beta f}^* (\mathbf{p}_{\beta b})
\phi_{pf}^* (\mathbf{p}_{pb})
\phi_{nf}^* (\mathbf{p}_{na})
d\mathbf{p}_{\beta b}
d\mathbf{p}_{pf}
d\mathbf{p}_{na}
\times
\]
\[
e^{i (p_{\beta b}'+ p_{p b}' - p_{n b})\cdot (x_b-x_a)} 
\langle \mathbf{p}_{\beta b}', \mathbf{p}_{pb}', \mathbf{p}_{na}'   
\vert S \vert \mathbf{p}_{\alpha a} \mathbf{p}_{pa} ,
\mathbf{p}_{nb}  \rangle
\times 
\]
\[
d\mathbf{p}_{\alpha a}
d\mathbf{p}_{pa}
d\mathbf{p}_{nb}
\phi_{\alpha i} (\mathbf{p}_{\alpha a})
\phi_{pi} (\mathbf{p}_{pa})
\phi_{ni} (\mathbf{p}_{nb})
 =
\]
\[
-2 \pi i 
\int
\phi_{\beta f}^* (\mathbf{p}_{\beta b})
\phi_{pf}^* (\mathbf{p}_{pb})
\phi_{nf}^* (\mathbf{p}_{na})
d\mathbf{p}_{\beta b}
d\mathbf{p}_{pf}
d\mathbf{p}_{na}
\times
\]
\[
\delta^4 ( p_{\alpha a} + p_{pa} + p_{nb} - p_{\beta b}' - p_{pb}' - p_{na}'
) e^{i (p_{\beta b}' + p_{pb}'- p_{nb}) (x_b-x_a)}
\times
\]
\[
\sum_j  {
\langle  (\mathbf{p}_{\beta b}',\mathbf{p}_{pb}')^+ \Vert 
V^{\dagger}_{j\beta} \Vert \mathbf{p}_{nb} , 
\mathbf{p}_{\nu_j} \rangle
\langle \mathbf{p}_{na}' ,
\mathbf{p}_{\nu_j} \Vert V_{j\alpha} \Vert 
(\mathbf{p}_{\alpha a},  \mathbf{p}_{pa})^- \rangle 
\over E_{\alpha a} + E_{pa} - E_{na}' - E_{\nu_j} + i0^+}  
\times
\]
\begin{equation} 
d\mathbf{p}_{\alpha a}
d\mathbf{p}_{pa}
d\mathbf{p}_{nb}
\phi_{\alpha i} (\mathbf{p}_{\alpha a})
\phi_{pi} (\mathbf{p}_{pa})
\phi_{ni} (\mathbf{p}_{nb})
 + \cdots 
\label{eq:c.10}.
\end{equation}

Equation (\ref{eq:c.10}) assumes that the weak interaction can be
treated using perturbation theory.  It was derived in a representation
in which the kinematic subgroup is the three-dimensional Euclidean
group.  Equivalent calculations using different kinematic subgroups
will have a similar form, but will have different ``off-shell''
neutrino variables.  This equation provides the starting point 
for derivations of the oscillation formula.

\section{Essential Ingredients for Oscillations}

Equation~(\ref{eq:c.10}) describes the scattering matrix element for
production and absorption of a superposition of neutrino mass
eigenstates as a second-order perturbation in the weak interaction.
Oscillations require that this expression can be approximated by a sum
of non-vanishing neutrino mass combinations multiplied by different
phase factors.  We now examine the essential assumptions that are
needed to derive the standard simple oscillation formula in the
literature from Eq.~(\ref{eq:c.10}).

\subsection{(almost) ultrarelativistic}

Experiments focused on neutrino mixing lead to constraints upon the
squared differences among neutrino masses, with a scale much less than
that of the virtual neutrino momenta, which make the neutrinos
very relativistic.

An important property of very relativistic neutrinos is that neutrinos
with different masses propagate with approximately the same velocity,
near the speed of light.  This ensures that the wave packets
associated different mass neutrinos continue to overlap over
macroscopic distances where oscillations can be observed.

For very relativistic neutrinos the energy of a mass eigenstate with
three-momentum $p_j=|\p_j|$ can be approximated by:
\beq
E_j \approx p_j + {m_j^2\over 2 p_j}.
\label{ar:1}
\eeq
In a representation like Eq.~(\ref{eq:c.10}), where the interactions
are chosen to be kinematically translationally invariant, the 
neutrino momenta, $p_j$,
are the same for all $j$.  Oscillations are
sensitive to the difference of these energies for different mass
neutrinos.  Contributions from corrections beyond this expression are
very small in the relativistic limit.

The leading phase differences are thus proportional to $(m_i^2 -m_j^2)=
(m_i-m_j)(m_i+m_j)$.  This needs to be small enough so variations in the
phase are small over the scale of the size of the single particle wave
packets.

It is not clear whether neutrinos with larger mass differences would
lead to oscillations.  It has been shown, for example, that a
non-relativistic massive neutrino cannot oscillate against much
lighter partners~\cite{Giunti:1991}.

\subsection{interactions and factorization}

The operator $V_{j\alpha}$ in Eq.~(\ref{eq:c.10}) connects the leptonic
and hadronic spaces, but factors into separate components for each
space.  This is the case whether one uses the contact Fermi
interaction or a diagram with $W$ exchange.  The hadronic contribution
will involve vector and axial vector current matrix elements that are
measurable; they represent a separate factor in the $S$ matrix and
contribute to the normalization of $S$, but not the oscillation
formula.

The leptonic contribution for our example is a matrix element of the
charge current interaction of the form
$\langle \beta \| J_{V-A}^\mu \| \nu_j \rangle$.  This matrix depends
in principle on the masses of the neutrino $j$ and the charged lepton
$\beta$.  However, the leading contribution to the neutrino mass comes
from the oscillation phase described above, and so we can substitute
an approximate form for the matrix element by substituting $\nu_j \to
\nu_0$, where $\nu_0$ here and henceforth denotes a massless neutrino
of arbitrary flavor.  The lepton matrix element does, however, depend
upon the mass of the charged lepton.  Put another way, the response of
the detector to a specific charged lepton flavor depends upon the
kinematics of this matrix element.

In the absence of mixing, the weak interaction could be expressed as an
operator $V_\alpha$ for each lepton flavor $\alpha$.  With mixing, the
weak interaction couples each lepton flavor $\alpha$ to each of
the neutrino mass eigenstates $j$ via an operator $V_{j\alpha}$. 

These points permit us to factor the weak interaction in the following
schematic way:
\beq
V_{j\alpha} = U_{j\alpha}V_{0\alpha},
\label{ar:2}
\eeq 
where $U$ is a unitary mixing matrix associated with the lepton-neutrino 
side of the interaction, and $V_{0\alpha}$ depends upon the mass of
the charged lepton but employs a massless neutrino of flavor
$\alpha$.  This approximation places all of the neutrino mass and mixing information
into the matrix $U$.

Different representations of the Poincar\'e group for this process
will in general involve different $V_\alpha$ that have different
kinematic factors.  However, the ability to factorize the interaction
via Eq.~(\ref{ar:2}) will lead to the same $U_{j\alpha}$.  That is, the
interaction could have a representation-dependent form, but very small
neutrino masses lead to the ability to factorize and thereby extract a
mixing matrix independent of the representation~\cite{Giunti:1993}.

To the extent that particle kinematics can deviate from the mass or
energy shell, contributions from the interactions can enter.  This
means that in general, approaches using different forms of dynamics
will yield different results at the level of perturbation theory,
although the exact expression to all orders will agree.  For the
problem at hand, off-shell neutrino kinematics corresponds to higher
order weak contributions, which are very small in magnitude; as noted
below, the deviations from mass/energy shell kinematics are also very
small.

Henceforth we also omit the $\pm$ labels in Eq.~\ref{eq:c.10} and use
plane-wave lepton and hadron states.

\subsection{delta functions and packets}

Equation~(\ref{eq:c.10}) reflects the use of time-ordered perturbation
theory, in which the intermediate neutrinos are on their mass shells
and the intermediate state can have an energy different from that of
the external initial or final state.  The energy denominator in 
Eq.~(\ref{eq:c.10}) can be
written as
\beq
{1 \over E_{\alpha a} + E_{pa} - E_{na}' - E_{\nu_j} + i0^+} =
-i\pi \delta(E_{\alpha a} + E_{pa} - E_{na}' - E_{\nu_j})
+ {\cal P}{1 \over E_{\alpha a} + E_{pa} - E_{na}' - E_{\nu_j}}.
\label{denom}
\eeq 
The energies of the initial and final particles that appear in this
denominator depend on the three-momenta of these particles.  When these
momenta are integrated over the factors in the numerator of
Eq.~(\ref{eq:c.10}), the delta function term only gives a contribution
for momenta that give an on-shell neutrino energy,  which leads to the
phase factor $\exp (-iE_\nu T)$, where the time difference $T=t_b-t_a$ is the
macroscopic time difference used to construct the asymptotic states,
which is very large on microscopic scales.

The principal-value term depends upon an integral that involves the
phase factor $\exp{(-iE T)}$ where $T$ is large and $E$ depends on the
momenta.  It is not hard to show that if the remaining part of the
integrand is a smooth function of $E$, then as $T$ becomes large the
principal value contribution will be exponentially suppressed relative
to the delta function in Eq.~(\ref{denom}).  In practice the integrand
includes Jacobians involving square roots and wave packets of unknown
smoothness.  Less restrictive assumptions on the smoothness of the
integrand lead to an algebraic suppression of the principal value
term for large $T$, but the contribution is still very small in
practice.  The general case was carefully examined by Grimus and
Stockinger \cite{Grimus:1996}.


The dominance of the energy delta function in Eq.~(\ref{denom}) means
that all four components of the neutrino four-momentum are
constrained:
\begin{equation}  
{p}^\mu_{\nu_j} =
{p}^\mu_{\alpha a}-  {p}^\mu_{pa} - {p}^\mu_{na} .
\label{4-momentum}
\end{equation} 
If we use this result to evaluate Eq.~(\ref{eq:c.10}) using plane-wave
momentum states, then there is no oscillation.  To see this, we note
that oscillations require the superposition of two or more neutrino
mass eigenstates.  Any particular combination of external particle
four-momenta will yield {\it at most one} neutrino four-momentum
$p_{\nu_j}$ satisfying Eq.~(\ref{4-momentum})~\cite{Cohen:2008qb,
  Akhmedov:2009rb}, and satisfying the mass condition $p_{\nu_j}^2 =
m_j^2$ of only one mass eigenstate.  For plane-wave external particle
states, there can be no superposition of neutrino mass eigenstates
that would set up an interference.

Thus, the oscillation effect necessarily entails momentum
distributions, or packets, of the external particles.  Under such
distributions, Eq.~(\ref{eq:c.10})  will in general give rise to a
superposition of neutrino mass eigenstates, with each term carrying
slightly different four-momenta.  Since the energy and three-momentum
values will vary inside the external wave packets, so too the energy
and three-momentum components of the neutrino mass eigenstates will
differ from each other.  The most that we can say is that the
contributions from different mass eigenstates will have quantitatively
{\it similar} components of their four-momenta as dictated by the
momentum ranges of the external wave packets.

It is important to note here that the term {\it packet} refers to the
momentum distributions of the external particles.  The neutrino mass
eigenstates also have a momentum distribution that will depend upon
the external packet shapes via the four-momentum condition of
Eq.~\ref{4-momentum}, but this distribution is not a wave packet in
the usual sense of preparation of observable states.

We have seen that external wave packets are essential to obtaining an
oscillation formula.  At the same time, the relevant distributions
must be narrow enough that the phase factor in Eq.~(\ref{eq:c.10}) can be
factored out of the integrals over these distributions.  When the
conditions for oscillation are satisfied, the phase is approximately
stationary.  The external distributions must therefore be narrow
enough to maintain this property.

A somewhat curious conclusion to this train of reasoning is that, once
the phases in the oscillation formula are fixed by the kinematics, they
depend, to lowest order, only on the neutrino eigenstate masses and
the {\it average} neutrino energy as determined by the external
particle kinematics.  One can therefore make the following
replacement:
\begin{equation}
\delta(E_{\alpha a} + E_{pa} - E_{na}' - E_{\nu_j}) \to
\delta(E_{\alpha a} + E_{pa} - E_{na}' - E_{\nu_0}),
\label{eq:deltareplacement}
\end{equation}
where, as noted above, $\nu_0$ denotes a massless neutrino of
arbitrary flavor.

\subsection{a matter of time}
The expression~(\ref{eq:c.10}) depends upon the space-time separation
between the points ($\mathbf{x}_a$, $t_a$) and ($\mathbf{x}_b$,$t_b$).
These points could be determined by events that can in principle be
measured.  However, in a typical oscillation experiment (e.g. a
reactor and a detector located a few kilometers away), the spatial
components $\mathbf{x}_a$ and $\mathbf{x}_b$ (and thereby the
displacement $\mathbf{L}$) are essentially known (at least to within a
scale characteristic of the reactor size), but the time difference
$T=t_b -t_a$ is not: the detector could record the time of a neutrino
detection event, but there is no corresponding record of the initial
weak interaction event.  Thus, the detector in principle is sensitive
to a range of possible times $T$.  As we shall see, the range of
values is in practice quite restricted, to the point where it is
possible to use any single value of $T$ within this range and obtain
the usual oscillation formula.

We now examine the variation of the $S$ matrix with the time $T$ in
Eq.~(\ref{eq:c.10}).  Equivalently, we can consider the variation as a
function of a velocity $v = |\mathbf{L}|/T$.  Since the neutrino mass
eigenstates are almost ultrarelativistic, we could first assume that $v=c=1$.
Alternatively, we could assume that $v$ corresponds to the velocity of
one of the neutrinos:
\beq
\mathbf{v} = \mathbf{v}_j = {\p_j\over E_j}.
\eeq
These seem to be reasonable assumptions, given the kinematics. 
For oscillations it should not matter which assumption is made.
Furthermore, conditions should be such that the oscillating phase 
approximately factors out the integrals in (\ref{eq:c.10}).

To study these conditions
consider the interference phase $\phi_{12}$ from mass eigenstates 1
and 2:
\beq
\phi_{12} = (p_1 - p_2) \cdot x
= L\left[{E_1-E_2\over v} -
(|\p_1|-|\p_2|)\right],
\label{eq:p.1}
\eeq
where the time $T$ has an associated velocity $v = L/T$.
If the neutrino masses $m_i$ are small compared to their energies, then
\beq
E_1 \approx |\p_1| + {m_1^2\over 2|\p_1|};\quad 
E_2 \approx |\p_2| + {m_2^2\over 2|\p_2|},
\label{eq:p.4}
\eeq

Now define $p,\delp$ such that $|\p_{1,2}| = p \pm \delp$. The
momentum difference $\delp$ has a maximum value determined by the
momentum distributions of the external particles.
We also express the velocity parameter $v$ in terms of an average
velocity ${\bar v}$ as $v={\bar v} + \delta v$, where
\beq
\bar{\mathbf{v}} = {1\over2}\left( {\p_1\over E_1} + {\p_2 \over E_2}\right);\quad
{\bar v} \approx 1 - {m_1^2\over 4 |\p_1|^2} - {m_2^2\over 4
  |\p_2|^2}.
\label{eq:p.5}
\eeq
The phase is then
\beq
\phi_{12} \approx
L\left[{m_1^2\over 2|\p_1|} - {m_2^2\over 2|\p_2|}
+(|\p_1|-|\p_2|) \left({m_1^2\over 4 |\p_1|^2} + {m_2^2\over 4
|\p_2|^2} - {\delta v}\right)\right].
\label{eq:p.6}
\eeq
Now define $p,\delp$ such that $|\p_{1,2}| = p \pm \delp$.  If we
further assume that $\delp\ll p$, then
\beq
\phi_{12} \approx
L{m_1^2-m_2^2\over 2 {\bar p}} - 2{\delta p}{\delta v}
+ O({\delta p}^2).
\label{eq:p.9}
\eeq
For the specific case that ${\delta v} = 0$,we are left with 
\beq
\phi_{12} \approx
L{m_1^2-m_2^2\over 2 p} + O(\delp^2).
\label{eq:p.8}
\eeq
In principle, ${\delta v}$ is arbitrary (as is the time), but its
contribution to the phase is limited by a factor involving $\delp$.
In fact, ${\delta v}$ is further limited by the range of the $p_i$
values that determine it.  For example, if, at one extreme, $v = v_1$,
then
\beq
\phi_{12} \approx
L\left[{m_1^2-m_2^2\over 2 {\bar p}} + {(m_1^2-m_2^2){\delta p}\over 2 {\bar p}^2}\right]
+ O({\delta p}^2),
\label{eq:p.10}
\eeq
that is, it generates a phase contribution that is suppressed by 
order $\delp/p$.  Thus,
we find that $\delp \ll 1$ in order to have interfering neutrino mass
eigenstate contributions, and to fix the time $T$ to within
correction factors that can be neglected.

\section{Result and Discussion}

Combining all of the simplifying results of the previous section, we
find that the space-time phase factors and the mixing matrices factor
out of the integrals over external wave packets, leaving a scattering
matrix element of the form
\beq
S_{fi}(x_b,x_a) \approx
\sum_j e^{-i \bar{p}_{\nu_j} \cdot (x_b-x_a)}
U_{j\alpha} U^*_{j\beta} s_{fi}(\beta,\alpha,j) 
\label{eq:c.13}
\eeq
where
\[
s_{fi}(\beta,\alpha,j) = 
\]
\[
-2 \pi^2  
\int 
d\mathbf{p}_{af}
d\mathbf{p}_{bf}
\phi_{bf}^* (\mathbf{p}_{bf})
\phi_{af}^* (\mathbf{p}_{af})
\delta^4 ( p_a + p_b - p_a' - p_b' )
\times
\]
\[
\langle  \mathbf{p}_{\beta b}',\mathbf{p}_{pb}' \Vert 
V^{\dagger}_{0\beta} \Vert \mathbf{p}_{nb} , 
\mathbf{p}_{\nu_0} \rangle
\delta ( E_{a} - E_{a'} - E_{\nu_0})  
\langle \mathbf{p}_{na}' ,
\mathbf{p}_{\nu_0} \Vert V_{0\alpha} \Vert 
(\mathbf{p}_{\alpha a},  \mathbf{p}_{pa}) \rangle \times
\]
\begin{equation}
d\mathbf{p}_{ai} d\mathbf{p}_{bi} 
\phi_{bi} (\mathbf{p}_{bi})
\phi_{ai} (\mathbf{p}_{ai}).
\label{eq:c.14}
\end{equation}

Integrating over wave packets satisfying the conditions described
above, we obtain a measurement probability for neutrino production via
initial flavor $\alpha$ and neutrino absorption via final flavor $\beta$:   
\begin{equation}
P^{\rm rel}_{\beta\alpha} = {\vert S_{fi}(x_b,x_a) \vert^2
 \over \vert s_{fi}\vert^2}  \approx
\sum_{jk} U^*_{j\beta} e^{-i \bar{p}_{\nu_j} \cdot (x_b-x_a)} U_{j\alpha}
U^*_{k\alpha} e^{- i \bar{p}_{\nu_k} \cdot (x_b-x_a)} U_{k\alpha} =
\eeq
\beq
\delta_{\beta\alpha} + 2 \Re \sum_{j > k} 
U^*_{j\beta} U_{j\alpha} U^*_{k\alpha} U_{k\beta} \exp
\left(i {\Delta m_{jk}^2 L \over 2 E_\nu} \right)
\label{eq:c.15}
\end{equation}
where 
\beq
\Delta m_{jk}^2 = m_j^2 - m_k^2.
\label{eq:c.16}
\eeq
Equation~(\ref{eq:c.15}) is the standard oscillation formula found in
the literature.  Its simplicity rests upon all of the elements
described in the previous section.

We conclude this section with a brief discussion of what is observable
among the elements of neutrino oscillations.

In principle, the mixing matrix $U$ is really part of the interaction
Hamiltonian, whose matrix elements describe the connection of charged
leptons (identified by flavor) with neutrino mass eigenstates, since
it is the latter (rather than neutrino flavor eigenstates) that must
form the basis of a the Poincar\'e group with the correct
transformation properties.  An interaction Hamiltonian in general is
not observable, because it is always possible to generate scattering
equivalent Hamiltonians that leave the observables unchanged.  One can
constrain the problem by completely specifying the Hilbert space, but
the extraction will depend upon this specification.  That said, all of
this uncertainty is eliminated if the interaction can be treated
perturbatively, as it can for many electromagnetic and weak
processes. One can still perform interaction-dependent unitary
transformations that lead to different Hamiltonians, but the effect on
extracting parameters such as mixing angles will always be of higher
order.

A perturbative $S$-matrix calculation that exhibits dependence upon
the choice of kinematic symmetry group in relativistic dynamics is
incomplete: the full calculation will be unitarily equivalent to other
full calculations based upon other symmetry groups.  We note here that
our derivation of the oscillation formula makes use of the symmetry of
the three-dimensional Euclidean group.  This leads to delta functions
in three-momenta in the interaction matrix elements.  However, the
three-momentum delta functions end up being paired (modulo very small
corrections) with an energy delta function.  As noted above, what
matters in the end is the resulting {\it four-momentum} delta function
together with external wave packets that determine the kinematics of
the neutrino mass eigenstates.  One could also formulate the
derivation using front-form dynamics, which exhibit the symmetry of
the null plane.  The interaction matrix elements contain a delta
function in front-form three-momenta ($\p_\perp, p^+ = p^0 + p^3$).
In turn, it is paired with a denominator in $p^- = p^0 - p^3$, which,
for the geometry in question, is again approximately a $p^-$ delta
function.  This again leads to a four-momentum delta function,
modulated by external wave packets.

\section{Conclusion}

In this paper we used an $S$-matrix approach to study the problem of
neutrino oscillations.  This approach is exact and only
involves initial and final states that can in principle be prepared
and detected in laboratory experiments.  From
this exact formalism we investigated the features that are
combined to arrive at the standard neutrino oscillation formula,
Eq.~(\ref{eq:c.15}).  We summarize these briefly below:

\begin{itemize} 

\item[1.] The perturbative nature of the weak interaction means that
  the oscillation formula can be derived from a second-order
  expression via the two-potential equation (\ref{eq:c.10}).

\item[2.] All three neutrino masses are small.  The leading
  contribution to the oscillation formula comes from the first
  correction to ultrarelativistic kinematics in the space-time phase.
  All remaining dependence upon neutrino mass can be neglected.

\item[3.] Wave packets are essential for oscillations.  The spread of
  external particle four-momenta makes it possible for all
  intermediate mass eigenstates to contribute coherently to the $S$
  matrix with slightly differeing four-momenta.

\item[4.] Explicit off-shell effects can be neglected.  Oscillations
  can be computed in a variety of approaches that differ by unitary
  transformations that depend upon the interaction.  These approaches
  will all have the same leading contribution to the oscillation
  formula, with all differences (which can be written as off-shell
  effects) being of higher order in the weak interaction.

\item[5.] The weak Hamiltonians in the second-order calculation can be
  factored into a product of terms, one containing only mixing
  information and the other describing a weak interaction involving a
  fictitious massless, flavored neutrino.  Corrections to
  factorization are of higher order in the interaction and/or the
  neutrino mass.

\item[6.] The oscillation formula depends upon the space-time
  separation between production and disappearance of the neutrino mass
  eigenstates.  The spatial separation is known, but the time is not
  measured.  Nevertheless, for reasonably small sizes of the external
  four-momentum distributions (e.g. detector resolution), the range of
  relevant times is restricted to values that produce the standard
  oscillation formula.

\end{itemize}

We conclude that the existence of a simple neutrino oscillation
formula that can be derived from a variety of theoretical perspectives
represents a remarkable convergence of several specific physical
properties of neutrinos and their interactions.  However, to improve
on this simple formula will in principle bring back all of the
interrelated effects that have so far been neglected, and
``corrected'' oscillation formulae may acquire theoretical
dependencies that the lowest-order expression does not carry.

The authors would like to express our gratitude to Boris Kayser for
helpful correspondence related to this manuscript.  One of the authors
(W.P.) would like to acknowledge support for this work by the U.~S.
Department of Energy, Office of Nuclear Physics, under contract
No. DE-FG02-86ER40286.

\bibliography{n-osc-ms-16}

\end{document}